# Special Issue on Collaborative Aspects of Open Data in Software Engineering

**J. Linåker**
RISE Research Institutes of Sweden

**P. Runeson**
Lund University

**A. Zuiderwijk**
Delft University of Technology

**A. Brock**
OpenUK

*Abstract*—High-quality data has become increasingly important to software engineers in designing and implementing today's software, for example, as an input to machine-learning algorithms and visualisation- and analytics-based features. Open data – i.e., data shared under a licence that gives users the right to study, process, and distribute the data to anyone and for any purpose – offers a mechanism to address this need. Data may originate from multiple sources, whether crowdsourced, shared by government agencies, or shared between commercial entities, and is undoubtedly inherent to all business and revenue models across the public sector, business and industry today. In this guest editorial for the Special Issue on Collaborative Aspects of Open Data in Software Engineering, we explore the collaborative aspects of open data in software engineering. We highlight how these aspects can benefit organisations, what challenges may exist and how these may be addressed based on current practice, and introduce the four papers included in this special issue.

■ **SOFTWARE ENGINEERS** require high-quality data for the design and implementation of today's software, especially in the context of Machine Learning (ML). This puts an emphasis on the need for publication and sharing of data from and between organisations, public as well as private. Following on from the paradigm of open innovation, open data provides a mechanism to increase the availability of such data, offering its utility and improving innovation and user choice through the inevitable interoperability this enables.

Consistent with previous work, this editorial defines open data as machine-readable data pro-actively shared on the internet under a licence that gives users the right to use, process, and distribute the data to anyone and for any purpose [1].

The open in open data highlights the potential for open innovation and collaboration around the data. Such collaboration can range from the sharing of new and connected data sets (including meta-data), to quality assurance, or enrichment and life cycle management of the data. The collaboration facilitates related standards and formats for the sharing and use of the data, as well as extending to the platform (i.e., software and infrastructure) used to enable the collaboration itself and the collection and use of the data (commonly in the form of Open Source Software (OSS)).

However, collaboration through openly sharing datasets is also complex and challenging. For example, it may be unclear which datasets are used for software and application development. Furthermore, open government data has barely been studied from a software development perspective. As a consequence, there is a lack of insight into the requirements of the software development community and how they leverage open data [2], although such insights have started to emerge in the literature [3].

This special issue of IEEE Software focuses on the collaborative aspects of open data in software engineering. This editorial is structured as follows. First, we explore the various collaborative aspects of open data in software engineering. Second, we highlight how these collaborative aspects can benefit organisations, and also what challenges may exist and how these may be addressed. Third, we provide an overview of the included papers in this special issue, and finally we give a short summary of practice.

COLLABORATIVE ASPECTS OF OPEN DATA IN SOFTWARE ENGINEERING

Collaboration on open data often occurs in what is called Open Data Ecosystems (ODE), where OpenStreetMap and Wikidata are two popular examples. These ecosystems represent a form of networked communities of actors (organizations and individuals), which base their relations to each other on a common interest [4]. The common interest is creating and providing free geographic data in the case of OpenStreetMap, and providing a free knowledge base that can be read and edited by both humans and machines, in the case of Wikidata.

An ODE is commonly supported by an underpinning technological platform that enables actors within the ecosystem to process data (e.g., find, archive, publish, consume, or reuse) as well as to foster innovation, create value, or support new businesses. Actors collaborate on the data and boundary resources (e.g., software and standards), through the exchange of information, resources, and artifacts. In the cases of OpenStreetMap and Wikidata, their respective platforms are openly available, including boundary resources, such as software for publishing and managing the data.

Another essential characteristic of an ODE is how its governance is established, where we differentiate between organization-centric, consortium-based, and community-based ODEs [4]. The aforementioned examples of OpenStreetMap and Wikidata are community-based, where the governance is spread out among the individuals who are members of the ecosystem. In organization-centric and consortium-based ODEs, governance is concentrated to a single or set group of actors, commonly public or private organizations with shared business interests in the data shared within the ecosystem. Examples from research can be found within domains such as labor market, public transport, smart cities, and industry 4.0, where both public and private organizations hold and share governance in different constellations [3].

Several actors may be involved in the collaboration in an ODE, creating a value chain ranging from the data providers to the data users. Lindman et al. [5] identified five roles in this type of collaboration: open data publishers, data extractors and transformers, data analyzers, user-experience providers, and support-service providers. These are needed to get a fully functional pipeline from data to user service, and may be filled by actors within or across organizations. The data publisher is commonly a public entity as sharing of open data from private entities is not yet a common phenomena [4].

As with all aspects of digitalization, open data involves collaboration between actors with competence in the digital, and actors knowledgeable of the application domain. Successfully combining the digital competences with the application domain requires willingness and ability to cross cultural and language barriers. Further, as legal conditions for the use and spreading of data is foundational for data-driven software development, the ability to understand and communicate with jurists is essential.

BENEFITS OF COLLABORATION ON OPEN DATA

Collaboration on open data has the potential to generate value similar to OSS and other types of Open Innovation. This comes as an effect from tapping into the wisdom of the crowd and exploiting the potential workforce within and outside of an ODE.

From a cost-saving perspective, the potential external workforce may help in tasks such as providing, collecting, processing and publishing data [4]. As highlighted, an ODE can be resembled to a value chain where the raw material of data gets enriched and processed in a collaborative manner [5]. Reaching outside of an ODE, crowdsourcing as well as mass collaboration can be used to increase participation and collaboration of external individuals [6].

From a quality perspective, collaboration on the open data can, e.g., help to address and correct errors, but also add information to the data through annotations and other kinds of meta-data [4]. As a consequence, the quality of ML-training sets and software using the data will be improved. This applies also to Open Governmental Data [3], where users may help improve data quality for the public agencies.

From an innovation perspective, the potential increased access to high-quality data can further help to provide new and extended training sets for ML-based applications as well as feature sets to other forms of data-driven software (e.g., Google Maps). Innovation can further be accelerated as new use cases and markets may get extended or created for the ODE, or even trigger the creation of new ones. Furthermore, collaborating through open data may lower the related entry cost for actors aiming to utilize data to offer services, help to catalyze new entrepreneurial efforts, increase transparency and accountability, and transform incumbents and public organizations through improved decisions and services [7].

CHALLENGES AND WAYS OF IMPROVING COLLABORATION ON OPEN DATA

Sharing and collaborating on open data brings forward a number of challenges, both technical and process-oriented, as well as cultural and business-oriented [4]. Below, we highlight a few of these challenges that practitioners may need to consider within an ODE or when thinking of entering or creating one

Business and Competition Aspects

From a business perspective, an important challenge concerns the motivation of why a data set should be shared in the first place [3]. There needs to be business incentives that align with the company's business model. Practitioners hence need to understand aforementioned benefits and be able to contextualize these in their own environment and relate to relevant business goals.

The benefits further need to be nuanced and weighed against the potential costs and risks of releasing the data. Potential costs may be related to the data management life-cycle, i.e., the collection, processing, quality assurance, sharing, and distribution of the data. As with OSS, these costs as well as the potential benefits relate to the amount of collaboration that actually takes place. Hence, actors within an ODE must find ways for facilitating and orchestrating a sustainable collaboration and sharing of data.

A specific challenge for such collaboration is the notion of co-opetition [4], i.e., a space where competitors can collaborate with each other without being afraid of giving away or losing their competitive edge [8], which can be a significant obstacle for commercial data sharing. Researchers and companies may not be willing to openly share their data on novel software innovations and services since this reduces their own ability to commercially exploit them [9]. On the other hand, commodity data may be a basis for co-opetition, as for example the OpenStreetMap demonstrates.

To manage and enable such co-opetition, there may be a need for a neutral governance actor within the ecosystem that can mediate discussions, craft a common vision for the ecosystem, and help actors share data that everyone is comfortable with [3]. The latter, commonly referred to as selective revealing, can involve e.g., that only certain abstractions of data are shared.

Technical Aspects

The potential collaborational aspects also bring up a number of more technical challenges [4]. One such challenge concerns the collection of data and how to ensure its quality. Some scholars indicate that datasets of insufficient quality may be misinterpreted or misused [10], rather than improving the quality of software. Common domain models and standards for how the data is shared and used, as well as transparent processes and OSS tools for collection and enrichment of the data, contribute to address this challenge.

Introducing feedback-loops both within an ODE and towards the end-users is another means, which is specifically highlighted and addressed by Rudmark and Andersson in this special issue. Versioning the data is another practical aspect that needs consideration in order to enable decentralized collaboration, similar to what can be observed in OSS ecosystems. Worthington et al. explore this topic further in their contribution to this special issue.

Cultural, Organizational, and Legal Aspects

Cultural and organizational aspects is another set of challenges [4], e.g., aligning strategic and operational levels within an organization on what data to share. Individuals may have different views or understanding of the risks and benefits that sharing would imply. These challenges may also be manifested in the collection and collaboration of the data, as again, individuals of different backgrounds and cultures may have mis-aligned perspectives. This challenge is further explored by Thurney et al. in this special issue, who highlight the need for practitioners to educate each other based on their different domain knowledge and understanding.

Another set of challenges are related to legal conditions [4]. Organizations may be reluctant to share data due to uncertainty about liability and what licences may imply in practice. Risk of legal complications due to GDPR is a specific and common concern under European legislation. Companies, for example, collect and maintain considerable proprietary employee and customer data which they may be reluctant to share through collaborations [5]. Enabling individuals to gain control of how their data is shared (cf. MyData[1]) may be one way of addressing this rather difficult challenge, an area that is further explored by Alorwu et al. in their contribution to this special issue.

OVERVIEW OF THE SPECIAL ISSUE PAPERS

This special issue covers different topics related to collaborative aspects of open data in Software Engineering. Out of the 9 submitted papers, we selected 4 papers for this special issue. We applied a rigorous review process to each article, including a review by at least three

---

[1] https://mydata.org

expert reviewers. Below we summarize the papers.

The first article deals with the originators of data. Alorwu et al. elaborate on the contributors of crowdsourced data in the health domain. Open health data may be used for research and as inputs for digital health software solutions. They surveyed 80 participants, who previously donated health data to a decision support system, asking about their willingness to donate data for public use. They find that data donators, despite donating data as "open", wanted to influence for what, by whom, and where their data were used. The respondents voice limited trust in private stakeholders, such as pharmaceutical and insurance companies, and bring up privacy concerns with donating data. The authors conclude by connecting these concerns with the MyData initiative, providing mechanisms for donors to keep control of their data, still allowing it to be used for permitted purposes.

Next, Rudmark and Andersson focus on the quality of data and which role feedback loops may have to improve data quality. The feedback may be given by the data publishers themselves, or by data re-users. Further, the feedback may come from internal or external use of the data. With examples from public transport data, they present data dogfooding – data providers use their own data; external application monitoring – the data provider monitor how their data is used in other actors' applications; community curation – the actors work together on improving the data quality; and external quality proxy – letting an external actor check the data before publishing it. These different approaches are applicable depending on the characteristics of the data ecosystem.

The need for data users to understand what has changed, is the topic addressed by Worthington et al. They report on a case with open customs tariff data, and observe users' strategies to overcome lack of change information. Based on the observations, they outline three approaches to communicate changes in an open data set. 1) publish change information as a "sidecar", i.e. outside the core system as a separate information entity, 2) publish versions with change tracks as an HTTP API endpoint to be consumed by users, and 3) integrate the versioning into the database and allow users to query changes. Being technically quite standard solutions, the case demonstrates how important it is that open data is not only published, but also is set under version control.

In the fourth article, Thurnay et al. address the cross-discipline communication and collaboration needed to create an open data ecosystem. They created a database of legal documents in Austria and report their lessons learned from the collaboration between law experts and technology experts. They were surprised by how much implicit domain knowledge each category of expert had, and how much effort it took to bridge the gap. However, they took the role of teachers for each other, and gradually the technology experts became genuine experts in a narrow field of the law, while the legal experts were able to read Python source code implementations of the text processing.

SUMMARY OF PRACTICE

Demand for high-quality data is growing as are the costs and resources for collecting and managing it. As a response, open data offers a new arena for open collaboration and innovation on data similar to OSS. Yet, software engineering practices have not kept pace in terms of enabling such collaboration. In part, there is a need for the evolution of open data ecosystems, similar to open source communities, that can facilitate such collaboration and sharing of data between actors, independent of aligning or competing interests.

Challenges include technical as well as cultural, organization and legal aspects as explored by the four articles included in this special issue. Altogether, they clearly show that creating open data ecosystems is a socio-technical endeavour, and they provide practice-based recommendations to mitigate problems that might appear. The request for high-quality data for a multitude of data-driven applications will not decline, and open data crowdsourcing, quality assurance, version control, and collaboration will keep playing a central role for the future of software engineering.

ACKNOWLEDGMENT

We are incredibly grateful for the support we received from several individuals who made this special issue possible and successful, including the anonymous reviewers, authors, and IEEE Software's editorial staff.

**Dr. Johan Linåker** is a Senior Researcher at RISE Research Institutes of Sweden. His research interests include empirical software engineering research in industry and public sector in the context of open innovation. He is specifically interested in the areas of Open Source Software and Open Data and how these can serve as a source of open innovation and help create value for software-intensive organizations, independently and through an ecosystem context. Johan also holds an interest in requirements engineering and product management. Contact him at johan.linaker@ri.se.

**Dr. Per Runeson** Dr. Per Runeson is a professor of software engineering at Lund University, Sweden, and leader of its Software Engineering Research Group (SERG). His research interests include empirical research and collaboration with industry on software development and management methods. He is particularly interested in studies on testing, and the role of open source and open data in software engineering. He is the principal author of "Case study research in software engineering", has co-authored "Experimentation in software engineering", serves on the editorial board of IEEE Transactions on Software engineering, and Software Testing, Verification and Reliability, the advisory board of Empirical Software Engineering, and is a member of several program committees. Contact him at per.runeson@cs.lth.se.

**Dr. Anneke Zuijderwijk,** is an Assistant Professor at the Faculty of Technology, Policy and Management at Delft University of Technology in the Netherlands. Her research focuses on open data, and more specifically, on theory development concerning infrastructural and institutional arrangements that incentivize open data sharing and use behavior by governments, researchers, companies and citizens. Anneke obtained her PhD with distinction, received the international Digital Governance Junior Scholar Award, serves as the Editor-in-Chief for the e-Journal of e-Democracy and Open Government, and was ranked as one of the most influential open data researchers worldwide (Hossain, Dwivedi & Rana, 2016). Her research has been cited over 5,300 times. Contact her at a.m.g.zuiderwijk-vaneijk@tudelft.nl.

**Amanda Brock** is the CEO of the UK body for Open Technology, being open source software, open hardware and open data, OpenUK. Amanda has (amongst many things) previously been the Chair of the Open Source and Intellectual Property (IP)

Advisory Group of the United Nations Technology Innovation Labs, and General Counsel of Canonical, one of the world's biggest open source companies and the commercial sponsor of Ubuntu, setting up the global legal team and running this for 5 years. Contact her at amanda.brock@openuk.uk.